\documentstyle[12pt]{article}

\begin{document}

\pagestyle{plain}
\setcounter{page}{1}
\setcounter{footnote}{00}
\renewcommand{\thefootnote}{\alph{footnote}}

\baselineskip=18pt 
\def\doublespaced{\baselineskip=\normalbaselineskip\multiply
    \baselineskip by 150\divide\baselineskip by 100}

\def\su{$SU(2)_{\em l} \times SU(2)_h\times U(1)_Y$\,}
\def\uem{$U(1)_{\rm{em}}$\,}
\def\suu{$SU(2)\times U(1)_Y$\,}
\def\beq{\begin{equation}}
\def\enq{\end{equation}}
\def\ra{\rightarrow}
\def\D0{D\O~}
\def\ETslash{\not{\hbox{\kern-4pt $E_T$}}}
\def\pslash{\not{\hbox{\kern -1.5pt $p$}}}
\def\kslash{\not{\hbox{\kern -1.5pt $k$}}}
\def\aslash{\not{\hbox{\kern -1.5pt $a$}}}
\def\bslash{\not{\hbox{\kern -1.5pt $b$}}}
\def\Dslash{\not{\hbox{\kern -4pt $D$}}}
\def\wslash{\not{\hbox{\kern -4pt $\cal W$}}}
\def\zslash{\not{\hbox{\kern -4pt $\cal Z$}}}
%
%
\begin{titlepage}
\baselineskip=0.3in
\begin{flushright}
\hfill {\small 
PKU-TP-97-20\\
MSUHEP-70825\\
THU-TP-97-08\\}
{hep-ph/9709275}\\ 
\end{flushright}

\vspace{0.2in}
\begin{center}
{\Large\bf  Supersymmetric QCD Parity Nonconservation in \\
Top Quark Pairs at the Tevatron }\\
\vspace{.2in}
Chong Sheng Li$^{(a)}$,  C.--P. Yuan$^{(b)}$, Hong-Yi Zhou$^{(c)}$ \\
\vspace{.2in}
${(a)}$ Department of Physics, Peking University, Beijing 100871, China\\
${(b)}$ Department of Physics and Astronomy, Michigan State University,\\
     East Lansing, Michigan 48824, USA\\
${(c)}$ Institute of Modern Physics and Department of Physics,\\
	Tsinghua University, Beijing 100084, China\\

\end{center}
\vspace{.3in}

\begin{center}\begin{minipage}{5in}
\baselineskip=0.25in
\begin{center} ABSTRACT\end{center}

In the supersymmetry (SUSY) models, because of the mass difference between 
the left- and right-top squarks, the supersymmetric QCD interactions 
can generate parity violating effects in the 
production of $ t \bar t$ pairs. We show that 
SUSY QCD radiative corrections to the parity violating asymmetry 
in the production rates of the left- and right-handed top quarks via the 
$q \bar q \ra t \bar t$ process can reach about 3\% at the Fermilab Tevatron
with $\sqrt{s}=2$\,TeV. This could be observable with an integrated 
luminosity larger than $2\,{\rm fb}^{-1}$. We also show that in these models
the SUSY QCD radiative corrections to the $t \bar t$ production rate are small
unless gluinos are very light, of the order of 1\,GeV.

\end{minipage}\end{center}
\vspace{.5in}


\end{titlepage}
\newpage

\baselineskip=18pt 
\renewcommand{\thefootnote}{\arabic{footnote}}
\setcounter{footnote}{0}

\section{ Introduction }
\indent

In a recent paper \cite{ttew}, we studied the
parity violating asymmetry induced from 
the  supersymmetric electroweak (SUSY EW)
and Yukawa (SUSY Yukawa) corrections at the one loop level.
Two classes of supersymmetry (SUSY) models were considered:
the minimal supergravity (mSUGRA) models \cite{sugra}
and the minimal supersymmetric models (MSSM) with 
scenarios motivated by current data
\cite{hkmssm,kanelight}.
After sampling a range of values of SUSY parameters in  the
region that might give large contributions to 
the parity-violating asymmetry ${\cal A}$, and which are also 
consistent with either of the above two classes of models,
we found that the asymmetry ${\cal A}$ due to the one-loop
SUSY EW ($\alpha m_t^2/m_W^2$) and SUSY Yukawa corrections 
for the production process $q\bar{q}\to g\to t\bar{t}$
at the upgraded Tevatron is generally small,
less than a few percent. However, the sign can be 
either positive or negative depending on the values of 
the SUSY parameters. 
(The effect from the Standard Model (SM) 
weak corrections
to this asymmetry is typically less than a fraction of percent
~\cite{toppv,toppvtwo}.) 

In the supersymmetric Standard Model, some superparticles experience
not only the electroweak interaction but also 
the strong interaction. Although the
SM QCD interaction respects the discrete symmetries of 
charge conjugation (C) and parity (P), the SUSY QCD interactions for 
superparticles, in their mass eigenstates, need not be C and P
invariant.
(Needless to say, in the strong 
interaction eigenstates, the SUSY QCD interaction 
is C- and P-invariant.)
For either the mSUGRA or the MSSM models, the masses of 
the left-stop (the supersymmetric partner of the left-handed top quark) 
and the right-stop can be 
noticeably different due to the large mass of the top quark.
This is a general feature of the 
supersymmetry models in which the electroweak symmetry is broken 
spontaneously via radiative corrections.
Since both the left-stop and the right-stop
 contribute to the loop corrections for 
the $t \bar t$ pair production process $q \bar q, gg \ra t \bar t$, 
the different masses of the top-squarks will induce 
a parity violating asymmetry.
It is this effect that we shall study in this paper.
Because the $t \bar t$ pairs are produced predominantly via the 
QCD process $q \bar q \ra t \bar t$ at the Tevatron
(a ${\rm p}{\bar {\rm p}}$ collider with CM energy $\sqrt{s}=2\,$TeV), 
we shall concentrate on
the SUSY QCD corrections for the $q \bar q$ fusion process. 
 We show that the parity violating asymmetry 
$|{\cal A}|$ at the Tevatron induced by SUSY EW, SUSY Yukawa and SUSY 
QCD effects could add up to more than 3\% for some of the SUSY models.
The SM tree level cross section is $4.34$\,pb for a 176 GeV top quark.
The SM QCD correction increases the rate by about 60\% \cite{smqcd},
while the SM EW correction is less than a percent \cite{smew}.
It is expected that there are about 2300 fully reconstructed
$b$-tagged $t \bar t$ events in the lepton plus jets mode collected 
by the two experimental groups
with a 2 ${\rm fb}^{-1}$ integrated luminosity \cite{tev2000}.
This amounts to a signal at $\sim 90\%$ c.l. (confidence level)
with 2 ${\rm fb}^{-1}$, or $99\%$ c.l. with 10 ${\rm fb}^{-1}$. 
Thus, a study of ${\cal A}$ at the Tevatron could yield information
about the allowed range of SUSY model parameter space.

\section{ SUSY QCD Corrections and Parity Violation }
\indent

\begin{flushleft} 
{\bf I. Squark mixings} 
\end{flushleft}

In the MSSM the mass eigenstates $\tilde{q}_1$ and
$\tilde{q}_2$ of the squarks
are related to the (strong) current eigenstates $\tilde{q}_L$ 
and $\tilde{q}_R$ via the mixing angle $\theta_{\tilde{q}}$ by 
\begin{equation}
\tilde{q} _1 = \tilde{q} _L \cos \theta_{\tilde{q}} + 
\tilde{q} _R \sin \theta_{\tilde{q}},~~~
\tilde{q} _2 = -\tilde{q} _L \sin \theta_{\tilde{q}} +
 \tilde{q} _R \cos \theta_{\tilde{q}} .
\label{mixing}
\end{equation}  
For the top squarks, 
the mixing angle $\theta_{\tilde{t}}$ and the masses $m_{\tilde{t}_{1,2}}$
can be calculated by diagonalizing the following 
mass matrix \cite{hkmssm},
\begin{eqnarray}
\label{eqnum2}
& &  M^2_{\tilde{t}} =\left(
	   \begin{array} {ll}
	     M^2_{\tilde{t}_L} & m_tm_{LR} \\
	     m_tm_{LR} & M^2_{\tilde{t}_R} 
	   \end{array}
	 \right),    \nonumber \\
& & M^2_{\tilde{t}_L}=m^2_{\tilde{t}_L} + m^2_t+(\frac{1}{2}
  -\frac{2}{3}\sin^2\theta_W)\cos(2\beta)m_Z^2 , \nonumber \\
& &  M^2_{\tilde{t}_R}=m^2_{\tilde{t}_R} + m^2_t
  +\frac{2}{3}\sin^2\theta_W\cos(2\beta)m_Z^2 , \nonumber \\
& & m_{LR}=-\mu\cot\beta+A_t ,
\end{eqnarray}   
where $m^2_{\tilde{t}_L}$ and $~m^2_{\tilde{t}_R}$ 
are the soft SUSY breaking 
mass terms of the left-stop and the right-stop,
$\mu$ is the coefficient 
of the  $H_1 H_2$ mixing term in the superpotential, $A_t$    
is the parameter describing the strength of soft SUSY breaking trilinear 
scalar interaction $\tilde{t}_L\tilde{t}_R H_2$, and
$\tan\beta=v_2/v_1$ is the ratio of the vacuum expectation values 
of the two Higgs doublets. $\theta_W$ is the weak mixing angle, and $m_Z$ 
is the mass of the $Z$ boson. 

\begin{flushleft} 
{\bf II. Renormalized amplitudes  and the asymmetry} 
\end{flushleft}

The effects of parity nonconservation can appear as an asymmetry in the
invariant mass ($M_{t \bar t}$) distributions as well as in the integrated
cross sections ($\sigma$) for $t_L$ and $t_R$ production.
The integrated asymmetry, after integrating over a range of 
$M_{t \bar t}$, is defined by \cite{ttew}
\begin{eqnarray}
{\cal A}
  & \equiv &\frac{ N_{R} -N_{L} }{ N_{R} +N_{L} }
     = \frac{ \sigma_{R} -\sigma_{L} }{ \sigma_{R} +
\sigma_{L} },                            
\end{eqnarray}
where $N_R$ and $N_L$ are the numbers of right-handed and
left-handed top quarks. 
Thus, $\sigma_{R} = \sigma_{RL} +\sigma_{RR}$
($\sigma_{L} = \sigma_{LR} +\sigma_{LL}$) is the cross section
for producing a $t \bar t$ pair in which the top quark is right-
(left-) handed, and the 
top antiquark is either left- or right-handed.

 Some of the one loop scattering
amplitudes of $ q \bar q \ra t \bar t$ were already presented 
in Refs. \cite{rateone,ratetwo} for calculating the 
total production rates of $ t \bar t$ pairs. 
To calculate the parity violating asymmetry ${\cal A}$
in the  $ t \bar t$ system, 
additional renormalized amplitudes are needed.
In terms of the tree-level amplitude, $M_{0}$, and the next-to-leading
order SUSY QCD corrections, $\delta M$, the renormalized amplitudes at
the one-loop level can be written as
$M=M_0+\delta M$.
Denote the momenta of the initial and the final state
particles as $ q_l(p_4)\bar q_m(p_3)\to t_i(p_2)\bar t_j(p_1)$,
and the Dirac four-spinor as
$u_{i} \equiv u(p_{i})$ ($v_i \equiv v(p_i)$)
for particle (anti-particle) $i$.
Then, $M_0=ig_s^2(T^c_{ji}T^c_{lm})J_1\cdot J_2/\hat{s}$,
where $J_1^\mu=\bar v(p_3)\gamma^\mu u(p_4)$
and $J_2^\mu=\bar u(p_2)\gamma^\mu v(p_1)$;
$\hat s$ is the invariant mass of the $t \bar t$ pair;
$g_s$ and $T^c_{ij}$ are the gauge coupling and the 
generator of the group $SU(3)_c$, respectively. 

To calculate the parity violating asymmetry
induced by the SUSY QCD effects, we follow the method presented in 
Ref. \cite{toppol}, in which the asymmetry was calculated 
numerically using the helicity amplitude method.
To obtain the renormalized scattering amplitudes, we adopt the
dimensional regularization scheme to regulate 
the ultraviolet divergences and the on-mass-shell 
renormalization scheme \cite{onshell} to define the input parameters.
The SUSY QCD corrections to the scattering amplitudes 
arise from the vertex diagram,
the gluon self-energy and the box diagrams, as well as the
crossed-box diagrams. 
The renormalized amplitudes can be written as
\begin{eqnarray}
\delta M & =& \delta M^{v1}+\delta M^{v2}+\delta M^s+ \delta M^{DB}
              +\delta M^{CB},
\end{eqnarray}
where $\delta M^{v1}$ and $\delta M^{v2}$ are vertex corrections,
$\delta M^{s}$ is the self-energy correction, and $\delta M^{DB}$
and $\delta M^{CB}$ are the contributions from the box diagrams and
crossed-box diagrams, respectively.
The results for these separate contributions are,
\begin{eqnarray}                                   
\delta M^{v1}& =& ig_s^2 (T^c_{ji}T^c_{lm})
{\bar u}(p_2)
 [F_0^{v1}\cdot J_1+F_1^{v1}\rlap/J_1+\rlap/J_1\rlap/F_3^{v1}
 +\rlap/F_4^{v1}\rlap/J_1+\rlap/F_6^{v1}\cdot J_1         \nonumber\\
 & &+ (F_1^{Av1}\rlap/J_1+\rlap/J_1\rlap/F_3^{Av1}
 +\rlap/F_4^{Av1}\rlap/J_1+\rlap/F_6^{Av1}\cdot J_1)\gamma_5]v(p_1)/\hat{s},
\end{eqnarray}                  
\begin{eqnarray}                                   
\delta M^{v2}& =& ig_s^2 (T^c_{ji}T^c_{lm})
{\bar v}(p_3)
 (F_1^{v2}\rlap/J_2+\rlap/F_6^{v2}\cdot J_2)u(p_4)/\hat{s},
\end{eqnarray}                  
\begin{eqnarray}                     
\delta M^{s}&=& F_0^{s}M_0,
\end{eqnarray}
\begin{eqnarray}                      
\delta M^{DB}& =&ig_s^2\frac{7}{6}(T_{ji}^cT_{lm}^c)[
F_1^{DB}(\bar u_2u_4\bar v_3v_1-\bar u_2\gamma_5
u_4\bar v_3\gamma_5v_1)\nonumber \\
& &+ F_2^{DB\mu}(-\bar u_2u_4\bar v_3\gamma_\mu v_1
-\bar u_2\gamma_\mu u_4\bar v_3v_1
-\bar u_2\gamma_5u_4\bar v_3\gamma_\mu\gamma_5 v_1
+\bar u_2\gamma_\mu\gamma_5u_4\bar v_3\gamma_5 v_1)\nonumber \\
& &+ F_3^{DB\mu\nu}(
\bar u_2\gamma_\mu u_4\bar v_3\gamma_\nu v_1
+\bar u_2\gamma_\mu\gamma_5 u_4\bar v_3\gamma_\nu\gamma_5 v_1)
\nonumber \\
& &+ F_4^{DB}(\bar u_2\gamma_5u_4\bar v_3v_1-\bar u_2
u_4\bar v_3\gamma_5v_1)\nonumber \\
& &+ F_5^{DB\mu}(-\bar u_2\gamma_5 u_4\bar v_3\gamma_\mu v_1
-\bar u_2\gamma_\mu\gamma_5 u_4\bar v_3v_1
-\bar u_2u_4\bar v_3\gamma_\mu\gamma_5v_1
+\bar u_2\gamma_\mu u_4\bar v_3\gamma_5v_1)\nonumber \\
& &+ F_6^{DB\mu\nu}(
\bar u_2\gamma_\mu u_4\bar v_3\gamma_\nu\gamma_5 v_1
+\bar u_2\gamma_\mu\gamma_5 u_4\bar v_3\gamma_\nu v_1)],
\end{eqnarray}
\begin{eqnarray}                      
\delta M^{CB}& =&ig_s^2\frac{1}{3}(T_{ji}^cT_{lm}^c)[
F_1^{CB}(\bar u_2u_3\bar v_4v_1-\bar u_2\gamma_5
u_3\bar v_4\gamma_5v_1)\nonumber \\
& &+ F_2^{CB\mu}(-\bar u_2u_3\bar v_4\gamma_\mu v_1
-\bar u_2\gamma_\mu u_3\bar v_4v_1
-\bar u_2\gamma_5u_3\bar v_4\gamma_\mu\gamma_5 v_1
+\bar u_2\gamma_\mu\gamma_5 u_3\bar v_4\gamma_5 v_1)\nonumber \\
& &+ F_3^{CB\mu\nu}(
\bar u_2\gamma_\mu u_3\bar v_4\gamma_\nu v_1
+\bar u_2\gamma_\mu\gamma_5 u_3\bar v_4\gamma_\nu\gamma_5 v_1)\nonumber \\
& &+ F_4^{CB}(\bar u_2\gamma_5 u_3\bar v_4 v_1-\bar u_2
u_3\bar v_4\gamma_5 v_1)\nonumber \\
& &+ F_5^{CB\mu}(-\bar u_2\gamma_5 u_3\bar v_4\gamma_\mu v_1
-\bar u_2\gamma_\mu\gamma_5 u_3\bar v_4 v_1
-\bar u_2u_3\bar v_4\gamma_\mu\gamma_5 v_1
+\bar u_2\gamma_\mu u_3\bar v_4\gamma_5 v_1)\nonumber \\
& &+ F_6^{CB\mu\nu}(
\bar u_2\gamma_\mu u_3\bar v_4\gamma_\nu\gamma_5 v_1
+\bar u_2\gamma_\mu\gamma_5 u_3\bar v_4\gamma_\nu v_1)],
\end{eqnarray}
where, $\rlap/J_1 \equiv  {J_1}^\mu \gamma_\mu$, {\it etc.}, and
the explicit expressions of the form factors
$F_i^{vj}$, $F_i^{Avj}$, $F_0^s$, $F_i^{DB,CB}$, $F_i^{DB(\mu, \mu\nu)}$, 
and $F_i^{CB(\mu, \mu\nu)}$ are given in the Appendix.

\section{  Numerical Results and Conclusion} 
\indent

In this section, we give our numerical results for a 176\,GeV top
quark \cite{topdisc}.
To avoid numerical instabilities and to take into account
the fact that 
the decay products of the produced top quarks at large scattering angles 
are better distinguishable from the background, we impose a cut on
the transverse momentum ($p_T$) and the rapidity ($y$) of
the top quark and anti-quark: 
\begin{equation}                                                     
p_T>20\;{\rm GeV} \quad \, {\rm and} \, \quad |y|<2.5 .\\
\label{kincut}
\end{equation}
We note that the parity violating asymmetry ${\cal A}$
would be independent of the parton 
distribution functions (PDFs) if there were no kinematic
cuts imposed in the calculation.
Therefore, ${\cal A}$ should not be sensitive to the choice
of the PDFs. In this paper we use the 
MRSA$^\prime$ PDFs \cite{mrsa} and evaluate
both the strong coupling $\alpha_s$ 
and the PDFs at the scale 
$Q=\sqrt{\hat{s}}=M_{t \bar t}$.\footnote{
Using the CTEQ4M PDFs \cite{cteq4} gives a similar result.}

As discussed in the previous section, the SUSY parameters relevant
to our study are 
$m_{\tilde{t}_1}$, $m_{\tilde{t}_2}$, $\theta_{\tilde{t}}$ (or 
$m_{\tilde{t}_{L}},m_{\tilde{t}_{R}},m_{LR}$),
$m_{\tilde{b}_{R}}$, $m_{\tilde{q}_{L,R}}$, and $m_{\tilde{g}}$.
To simplify our discussion, we assume 
$m_{\tilde{q}_{L,R}} = m_{\tilde{b}_{R}} =  m_{\tilde{t}_{L}}$,
so that there are only four SUSY parameters to be considered,
$m_{\tilde{t}_1}$, $m_{\tilde{t}_2}$, $\theta_{\tilde{t}}$ and
 $m_{\tilde{g}}$.
(The $SU(2)_L$ gauge symmetry requires that
$m^2_{\tilde{b}_{L}}=m^2_{\tilde{t}_{L}}$.)
  
The mSUGRA models predict radiative breaking of the electroweak 
gauge symmetry induced by the large top quark mass. 
Consequently, it is possible to have large splitting in the masses of
the left-stop and the right-stop, while the masses of all the other 
(left- or right-) squarks are about the same \cite{kanesugra}.
For the MSSM models 
with scenarios motivated by current data \cite{kanelight},
a light $\tilde{t}_{1}$ is likely to be the right-stop ($\tilde t_R$), 
with a mass at the order of $m_W$; the other  
squarks are heavier than $\tilde{t}_{1}$.
Since heavy superparticles decouple in loop contributions, we expect 
that a lighter $\tilde{t}_{1}$ would induce a larger asymmetry. 
Because the parity-violating effects from the SUSY QCD interactions arise 
from the mass difference between $\tilde{t}_{1}$ and $\tilde{t}_{2}$,
it is obvious from Eq. (\ref{mixing}) that the largest parity violating
effect occurs when $\theta_{\tilde{t}}$ is $\pm \pi/2$
for $m_{\tilde{t}_{R}} \le m_{\tilde{t}_{L}}$.
When $\theta_{\tilde{t}}=\pm \pi/4$, the parity asymmetry should be zero.
This is evident from the results shown in the Appendix, which 
indicate that the amplitudes that contribute to ${\cal A}$ are all 
proportional to $Z_i=\mp \cos(2\theta_{\tilde{t}})$.

In either the mSUGRA or the MSSM models,
the gluinos are usually as heavy as the light squarks,
on the order of a few hundred GeV. However, 
Farrar has argued \cite{lightgluino} that light gluinos are
still a possibility. If gluinos are light, then a heavy top quark can decay 
into a stop and a light gluino for 
$m_{\tilde{t}_1} < ( m_t - m_{\tilde{g}} )$ such that 
the branching ratio of $ t \ra b W^+$
could show a 
large difference from that ($\sim 100\%$) predicted by the SM. 
The CDF collaboration has measured 
the branching ratio of $ t \ra b W^+$ to be 
$0.87{^{+0.13}_{-0.30}}{^{+0.13}_{-0.11}}~$~\cite{br}.
At the $1 \sigma$ level, this implies that  
a 50 (90) GeV $\tilde{t}_{1}$ requires the mass of the gluino to be
larger than about 120 (80) GeV for $\theta_{\tilde{t}}=\pm \pi/2$.
However, at the $2 \sigma$ level (i.e. 95\% c.l.),
there is no useful limit on the mass of the gluino.\footnote{
A recent analysis on the possible experimental signature for a light 
gluino from top quark decay can be found in Ref. \cite{lgluinofnal}.}

To represent different classes of SUSY models in which the 
parity-violating asymmetry induced by the SUSY QCD interactions can
be large, we show in Table 1 four representative sets of models.
They are labeled by the set of parameters 
$(m_{\tilde{t}_1},m_{\tilde{t}_2},\theta_{\tilde{t}})$, which are equal to 
$(50,1033,-1.38)$, $(90,1033,-1.38)$, $(50,558,-1.25)$
and $(90,558,-1.25)$, respectively. 
(All the masses are in units of GeV.)
Based upon Eq. (2), one can also label these models by
$(m_{\tilde{t}_{L}},m_{\tilde{t}_{R}},m_{LR})$, which are
$(1000,90,1100)$, $(1000,118,1100)$, $(500,40,520)$
and $(500,88,520)$, respectively, for $\beta=\pi/4$.
\begin{table}
\caption{ }
Parity violating asymmetry ${\cal A}$
in $p \bar p \ra t \bar t +X $, as a function of $m_{\tilde{g}}$,
for four sets of SUSY models labeled by 
$(m_{\tilde{t}_1},m_{\tilde{t}_2},\theta_{\tilde{t}})$.
\vspace{0.1in}
\small{
\begin{center}
\begin{tabular}{|c|c|c|c|c|}
\hline
$m_{\tilde{g}}$ (GeV) &$(50,1033,-1.38)$ & 
$(90,1033,-1.38)$ & $(50,558,-1.25)$ &
$(90,558,-1.25)$ \\ \hline
2   & -1.10\% & -1.28\% &-0.98\% &-1.13\%  \\
50  & -1.53\% & -1.77\% &-1.40\% &-1.59\%  \\
100 & -2.34\% & -2.23\% &-2.21\% &-2.23\%  \\
120 & -2.86\% & -1.99\% &-2.89\% &-1.99\%  \\
135 & -3.16\% & -1.79\% &-3.43\% &-1.82\%  \\
150 & -2.58\% & -1.50\% &-2.80\% &-1.53\%  \\
175 & -1.18\% & -0.44\% &-1.30\% &-0.56\%  \\
200 &  0.99\% &  0.97\% & 0.82\% & 0.77\%  \\
225 &  1.60\% &  1.41\% & 1.40\% & 1.19\%  \\
250 &  1.53\% &  1.34\% & 1.35\% & 1.17\%  \\
275 &  1.27\% &  1.16\% & 1.16\% & 1.02\%  \\
300 &  1.04\% &  0.94\% & 0.95\% & 0.83\%  \\
\hline
\end{tabular}
\end{center}
}
\end{table}

It is interesting to note that for all the models listed in
Table 1, the
asymmetry ${\cal A}$ is negative (i.e. $\sigma_R < \sigma_L$)
for $m_{\tilde{g}}< 200$\,GeV, and its magnitude can be as large as
3\% for models with light ${\tilde{t}_1}$. 
The maximal $|{\cal A}|$ occurs when $m_{\tilde{g}}$ is about equal 
to $(m_t - m_{\tilde{t}_1})$ because of the mass threshold enhancement.
For $m_{\tilde{g}} > 200$\,GeV, the asymmetry ${\cal A}$ becomes
positive, with a few percent in magnitude, 
and monotonically decreases as $m_{\tilde{g}}$ increases.
Comparing these results with those induced by the SUSY EW and SUSY Yukawa
corrections \cite{ttew}, it is clear that  SUSY QCD interactions can
generate a relatively larger parity-violating asymmetry. 

The differential asymmetry 
${\cal A}(M_{t\bar{t}})$ also exhibits an interesting behaviour
as a function of the $t \bar t$ invariant mass $M_{t\bar{t}}$.
This is illustrated in Table 2 for the first SUSY model in
Table 1 ($(m_{\tilde{t}_1},m_{\tilde{t}_2},\theta_{\tilde{t}})$ =
(50, 1033, -1.38)).
As shown,  $|{\cal A}(M_{t\bar{t}})|$ increases as
$M_{t\bar{t}}$ increases for $m_{\tilde{g}} < 200$\,GeV, which is 
similar to the effects from the SUSY EW and SUSY Yukawa 
contributions \cite{ttew}.
When $M_{t\bar{t}}$ is larger than 500\,GeV, 
$| {\cal A}(M_{t\bar{t}})|$ can become very large at the expense
of the signal event rate.
For the model with $m_{\tilde{g}}$ equal to 200\,GeV, 
$ {\cal A}(M_{t\bar{t}})$ is positive for 
$M_{t\bar{t}}$ less than $\sim 500$\,GeV; it
reaches its maximal value when $M_{t\bar{t}}$ is about 
twice the gluino mass. This is again due to the
mass threshold enhancement.

\begin{table}
\caption{ }
The differential asymmetry $ {\cal A}(M_{t\bar{t}})$ 
and cross section ${\rm d}\,\sigma /{\rm d}\, M_{t\bar{t}}$
(in unit of fb/GeV)
as a function of $M_{t\bar{t}}$ for the first SUSY model 
 in Table 1 with various $m_{\tilde{g}}$ values.
\vspace{0.1in}
\small{
\begin{center}
\begin{tabular}{|c|cc|cc|cc|}
\hline
$M_{t\bar{t}}$ (GeV) & $m_{\tilde{g}}=2$\,GeV &&
$m_{\tilde{g}}=120$\,GeV && $m_{\tilde{g}}=200$\,GeV &
\\ \hline
 358 & -0.22\%  & 23.0 & -0.73\%  & 16.3  & 0.44\%  & 19.4  \\ 
 368 & -0.42\%  & 36.6 & -1.33\%  & 26.2  & 0.95\%  & 31.4   \\ 
 378 & -0.53\%  & 40.2 & -1.63\%  & 29.0  & 1.39\%  & 34.9   \\ 
 388 & -0.67\%  & 38.7 & -2.00\%  & 28.1  & 2.12\%  & 34.2  \\ 
 398 & -0.74\%  & 37.2 & -2.17\%  & 27.2  & 3.33\%  & 34.4    \\ 
 408 & -0.82\%  & 35.0 & -2.38\%  & 25.7  & 3.67\%  & 32.4   \\ 
 425 & -0.93\%  & 30.9 & -2.64\%  & 22.8  & 2.35\%  & 27.0   \\ 
 450 & -1.13\%  & 24.1 & -3.07\%  & 18.0  & 1.05\%  & 20.0   \\ 
 475 & -1.31\%  & 18.3 & -3.40\%  & 13.8  & 0.11\%  & 14.7   \\ 
 500 & -1.47\%  & 13.7 & -3.68\%  & 10.4  & -0.62\% & 10.8    \\ 
 525 & -1.61\%  & 10.3 & -3.81\%  &  7.9  & -1.21\% &  8.0  \\ 
 550 & -1.76\%  &  7.7 &  -4.11\% &  5.9  &  -1.71\%&  5.9 \\ 
 575 & -1.93\%  &  5.7 &  -4.34\% &  4.4  & -2.08\% &  4.3 \\ 
 1000 & -3.26\% & 0.032 &  -5.82\%& 0.026  & -4.66\% &  0.024  \\ 
\hline
\end{tabular}
\end{center}
}
\end{table}

We note that our conclusion for the magnitude of the 
parity-violating asymmetry
is different from that given in Ref. \cite{uill}.\footnote{
Without the cuts in (\ref{kincut}), the values of ${\cal A}$ 
for the first model in Table 1 are
$-1.0\%$, $-2.65\%$, and $+0.94\%$ for $m_{\tilde{g}}=2,120,200$\,GeV, 
respectively.
}
There the box (and crossed-box) diagram contributions 
were not included and some of the formulae 
(Eq. (8) and the form factor $A$ in the Appendix) for
calculating the vertex diagram contributions
appear to contain misprints. 
From our calculation, we find 
that the box diagram contributions
are in general small compared to the vertex diagram contributions,
and are sizable only for heavy gluinos ($m_{\tilde{g}} > 200$\,GeV).
Therefore, the fact that we have included these corrections, while Ref. \cite{uill}
did not, does not constitute the main difference between the two
calculations.
To  resolve this discrepancy, we compare our results with
those in the literature for  the SUSY QCD radiative correction
to the production rate of $q \bar q \ra t \bar t$ at the Tevatron
\cite{hagiwara}.
After correcting the relative sign between the box and crossed-box 
terms and the apparent misprints in the 
form factors $F_{12,13}^{DB}$ and $F_{12,13}^{CB}$
in Ref. \cite{hagiwara},\footnote{
These apparent problems in Ref. \cite{hagiwara}
were also pointed out in Ref. \cite{uill}.}
our results agree with those presented in \cite{hagiwara},
providing us with confidence in our results.
However, our numerical results for the $t \bar t$ production rate
do not agree with those given in Ref. \cite{uill}.
For the case considered in \cite{uill}, $m_{\tilde{g}}=200$\,GeV and
$m_{\tilde{t}}=m_{\tilde{q}}=75$\,GeV, we obtain a $39\%$, in contrast to 
$33\%$, correction in the total cross section without cuts. Including 
cuts in (\ref{kincut}) only slightly increases
the correction to $40\%$.

For completeness,  in Table 3 we show the SUSY QCD corrections 
$\Delta \sigma$ to 
the $q \bar q \ra t \bar t$ production rates at the Tevatron with 
$\sqrt{s}=2$\,TeV, as a function of $m_{\tilde{g}}$, 
for the first SUSY model in Table 1.
\begin{table}
\caption{ }
The SUSY QCD corrections ($\Delta \sigma$) to the $q \bar q \ra t \bar t$ 
production rates at the Tevatron with $\sqrt{s}=2$\,TeV 
as a function of $m_{\tilde{g}}$, 
for the first SUSY model  in Table 1.
\vspace{0.1in}
\small{
\begin{center}
\begin{tabular}{|c|cccccccccccc|}
\hline
$m_{\tilde{g}}$ (GeV) &
2 & 50& 100& 120& 135& 150& 175& 200& 225& 250& 275& 300
\\ \hline
$\Delta \sigma$ (pb) &
1.17 & 0.26& -0.04& -0.18& -0.87& -0.49& -0.02& 0.33& 0.30& 
0.24& 0.19& 0.16
\\ \hline
\end{tabular}
\end{center}
}
\end{table}
The SM tree level cross section, with the cuts in (\ref{kincut}), 
is 4.06 pb.
At the upgraded Tevatron, it is expected to measure the cross section 
of $t \bar t$ pair production to $\sim 10\%$ ($6\%$)
with a 2 ${\rm fb}^{-1}$ (10 ${\rm fb}^{-1}$)
integrated luminosity \cite{tev2000}.
Therefore, it is a challenging task to 
detect the difference in rates from SUSY QCD contributions
predicted by these models unless gluinos are very light 
(of the order of 1\,GeV).
Nevertheless, the parity violating effect induced by the 
SUSY interactions could be observable 
with a large integrated luminosity ($\ge 2 \, {\rm fb}^{-1}$) 
at the Tevatron.

Up to now, we have only considered the one loop SUSY QCD effects on 
the parity violating asymmetry ${\cal A}$ in $t \bar t$ 
pair production. 
Amusingly, the parity-violating asymmetry induced by the
SUSY QCD interactions can also occur at the Born level.
If gluinos are very light, of the order of 1\,GeV, this
asymmetry can be generated by the tree level process 
$\tilde{g}\tilde{g} \ra t \bar t$.  Unfortunately, its
production rate is smaller
than the $g g \ra t \bar t$ rate, which is only about one tenth of the 
$q \bar q \ra t \bar t$ rate at the Tevatron. 
Hence, it cannot be measured at the Tevatron. 
However, at the CERN Large Hadron Collider (LHC), 
the production rate of $\tilde{g}\tilde{g} \ra t \bar t$ 
is large enough to allow the measurement of the parity-violating asymmetry 
induced by the SUSY QCD interactions. 
The asymmetry in the production rates of $t_L \bar t$ and $t_R \bar t$, 
generated by the $\tilde{g}\tilde{g}$ fusion process alone,
can reach about 10\% for $M_{t\bar{t}}$ larger than about 500\,GeV. 
We shall present its details and include the effect
from the $gg$ and $q \bar q$ fusion processes in a 
future publication \cite{gluinopdf}.

\vspace{.5cm}

This work is supported in part by the National Natural Science Foundation
of China, and by the U.S. NSF grant PHY-9507683.

\vspace{1cm}
\begin{center}
{\bf Appendix }
\end{center}

We give here the form factors for the matrix elements appearing in
Eqs. (8)-(12).  
They are written in terms of the conventional one-, two-, three- and 
four-point scalar loop integrals defined in Ref. ~\cite{refeleven}.  

\baselineskip=12pt 

\begin{eqnarray*}
F_0^{v1\mu}&= & \sum\limits_{i=1,2}\frac{3\alpha_s}{8\pi}(
2m_{\tilde{g}}Y_i)C^{\mu}
(-p_2,k,m_{\tilde{t}_i},m_{\tilde{g}},m_{\tilde{g}})\\
& &+ \sum\limits_{i=1,2}\frac{\alpha_s}{12\pi}[
m_{\tilde{g}}Y_i((p_2-p_1)^{\mu}C_0/2-C^{\mu}]
(-p_2,k,m_{\tilde{g}},m_{\tilde{t}_i},m_{\tilde{t}_i})\\
& & \\
F_1^{v1}&= & \sum\limits_{i=1,2}\frac{3\alpha_s}{8\pi}[
C^{20}X_i-(m_t^2+m^2_{\tilde{g}})C_0X_i-2m_tm_{\tilde{g}}C_0Y_i]
(-p_2,k,m_{\tilde{t}_i},m_{\tilde{g}},m_{\tilde{g}})\\
& &+ \sum\limits_{i=1,2}\frac{\alpha_s}{3\pi}[
B_1X_i-2m_tm_{\tilde{g}}B_0^\prime Y_i+2m_t^2 B_1^\prime X_i]
(m_t^2,m_{\tilde{g}},m_{\tilde{t}_i})\\
& & \\
F_3^{v1\mu}&=& \sum\limits_{i=1,2}\frac{3\alpha_s}{8\pi}\big(
m_tC^{\mu}X_i\big)
(-p_2,k,m_{\tilde{t}_i},m_{\tilde{g}},m_{\tilde{g}})\\
& & \\
F_4^{v1\mu}&=& F_3^{v1\mu}\\
& & \\
F_6^{v1\mu\nu}&= & \sum\limits_{i=1,2}\frac{3\alpha_s}{8\pi}(-2X_i)C^{\mu\nu}
(-p_2,k,m_{\tilde{t}_i},m_{\tilde{g}},m_{\tilde{g}})\\
& &+ \sum\limits_{i=1,2}\frac{\alpha_s}{12\pi}[
X_i((p_2-p_1)^{\nu}C^{\mu}/2-C^{\mu\nu})]
(-p_2,k,m_{\tilde{g}},m_{\tilde{t}_i},m_{\tilde{t}_i})\\
  & & \\
F_1^{Av1}&= & \sum\limits_{i=1,2}\frac{3\alpha_s}{8\pi}Z_i\big(
C^{20}+(m_t^2-m^2_{\tilde{g}})C_0\big)
(-p_2,k,m_{\tilde{t}_i},m_{\tilde{g}},m_{\tilde{g}})\\
& &+ \sum\limits_{i=1,2}\frac{\alpha_s}{3\pi}Z_iB_1
(m_t^2,m_{\tilde{g}},m_{\tilde{t}_i})\\
& & \\
F_3^{Av1\mu}&=& \sum\limits_{i=1,2}\frac{3\alpha_s}{8\pi}Z_i
m_tC^{\mu}(-p_2,k,m_{\tilde{t}_i},m_{\tilde{g}},m_{\tilde{g}})\\
& & \\
F_4^{Av1\mu}&=& -F_3^{Av1\mu}\\
& & \\
F_6^{Av1\mu\nu}&= & \sum\limits_{i=1,2}\frac{3\alpha_s}{8\pi}(-2Z_i)
C^{\mu\nu}(-p_2,k,m_{\tilde{t}_i},m_{\tilde{g}},m_{\tilde{g}})\\
& &+ \sum\limits_{i=1,2}\frac{\alpha_s}{12\pi}Z_i[
((p_2-p_1)^{\nu}C^{\mu}/2-C^{\mu\nu})]
(-p_2,k,m_{\tilde{g}},m_{\tilde{t}_i},m_{\tilde{t}_i})\\
  & & \\
F_1^{v2}&= & \frac{3\alpha_s}{4\pi}\big(
C^{20}-m^2_{\tilde{g}}C_0\big)
(p_4,-k,m_{\tilde{q}},m_{\tilde{g}},m_{\tilde{g}})\\
& &+ \frac{2\alpha_s}{3\pi}B_1
(m_q^2,m_{\tilde{g}},m_{\tilde{q}})\\
& & \\
F_6^{v2\mu\nu}&= & \frac{3\alpha_s}{4\pi}(
-2)C^{\mu\nu}(p_4,-k,m_{\tilde{q}},m_{\tilde{g}},m_{\tilde{g}})\\
& &+ \frac{\alpha_s}{6\pi}[
-(p_2-p_1)^{\nu}C^{\mu}/2-C^{\mu\nu}]
(p_4,-k,m_{\tilde{g}},m_{\tilde{q}},m_{\tilde{q}})\\
  & & \\
F_0^s&=&\frac{3\alpha_s}{2\pi}\bigg(
(B_{21}+B_1+\frac{1}{6}
-(m_{\tilde{g}}^2(B_0+1)-2B_{22})/k^2)(k^2,m_{\tilde{g}},m_{\tilde{g}})\\
& & -(B_{21}+B_1+\frac{1}{6}-m_{\tilde{g}}^2B_0^\prime+2B_{22}^\prime)
(0,m_{\tilde{g}},m_{\tilde{g}})
\bigg)\\
& & +\sum\limits_{\tilde{q}}\frac{\alpha_s}{4\pi}\bigg(
(2B_{22}(k^2,m_{\tilde{q}},m_{\tilde{q}})-A_0(m_{\tilde{q}}))/k^2
-2B_{22}^\prime(0,m_{\tilde{q}},m_{\tilde{q}})
\bigg)\\
& & \\
F_1^{DB}&=& \sum\limits_{i=1,2}\frac{\alpha_s}{4\pi}\big(
(m_{\tilde{g}}^2+m_t^2)X_i+2m_tm_{\tilde{g}}Y_i\big)
D_0(-p_2,p_4,p_3,m_{\tilde{t}_i},m_{\tilde{g}},m_{\tilde{q}},m_{\tilde{g}})\\
& &\\
F_2^{DB\mu}&=& \sum\limits_{i=1,2}\frac{\alpha_s}{4\pi}\big(
m_tX_i+m_{\tilde{g}}Y_i\big)D^{\mu}
(-p_2,p_4,p_3,m_{\tilde{t}_i},m_{\tilde{g}},m_{\tilde{q}},m_{\tilde{g}})\\
& &\\
F_3^{DB\mu\nu}& =& \sum\limits_{i=1,2}\frac{\alpha_s}{4\pi}X_iD^{\mu\nu}
(-p_2,p_4,p_3,m_{\tilde{t}_i},m_{\tilde{g}},m_{\tilde{q}},m_{\tilde{g}})\\
& &\\
F_4^{DB}&=& \sum\limits_{i=1,2}\frac{\alpha_s}{4\pi}Z_i\big(
m_t^2-m_{\tilde{g}}^2\big)
D_0(-p_2,p_4,p_3,m_{\tilde{t}_i},m_{\tilde{g}},m_{\tilde{q}},m_{\tilde{g}})\\
& &\\
F_5^{DB\mu}&=& \sum\limits_{i=1,2}\frac{\alpha_s}{4\pi}Z_im_tD^{\mu}
(-p_2,p_4,p_3,m_{\tilde{t}_i},m_{\tilde{g}},m_{\tilde{q}},m_{\tilde{g}})\\
& &\\
F_6^{DB\mu\nu}& =&\sum\limits_{i=1,2}\frac{\alpha_s}{4\pi}Z_i D^{\mu\nu}
(-p_2,p_4,p_3,m_{\tilde{t}_i},m_{\tilde{g}},m_{\tilde{q}},m_{\tilde{g}})\\
& &\\
F_{1,3,4,6}^{CB}&=&F_{1,3,4,6}^{DB}(p_2\to p_1)\\
F_{2,5}^{CB}&=&-F_{2,5}^{DB}(p_2\to p_1)\\ .
\end{eqnarray*} 
\noindent\noindent
In the above:
\begin{eqnarray*}
& & k=p_1+p_2=p_3+p_4,\;\;\hat{s}=k^2,\;\;\\
& & X_i=a_i^2+b_i^2=1,~~~Y_i=a_i^2-b_i^2=\mp\sin(2\theta_{\tilde{t}}),
~~~Z_i=2a_ib_i=\mp\cos(2\theta_{\tilde{t}}).
\end{eqnarray*} 
where  $a_1,~b_1,~a_2,~b_2$ are given by
\begin{eqnarray*}
& & a_1=\frac{1}{\sqrt{2}}(\cos\theta_{\tilde{t}}-\sin\theta_{\tilde{t}}),\;\; 
 b_1=-\frac{1}{\sqrt{2}}(\cos\theta_{\tilde{t}}+\sin\theta_{\tilde{t}}),
\nonumber\\
& & a_2=-\frac{1}{\sqrt{2}}(\cos\theta_{\tilde{t}}+
\sin\theta_{\tilde{t}})=b_1,\;\;
 b_2=-\frac{1}{\sqrt{2}}(\cos\theta_{\tilde{t}}-\sin\theta_{\tilde{t}})=-a_1.
\end{eqnarray*}  
\noindent\noindent
Also,
\begin{eqnarray*}
& & B_0^\prime=\frac{\partial B_0(p^2,m_1,m_2)}{\partial p^2}, 
~~B_1^\prime=\frac{\partial B_1(p^2,m_1,m_2)}{\partial p^2}, 
~~B_{22}^\prime=\frac{\partial B_{22}(p^2,m_1,m_2)}{\partial p^2}, \\ 
& &  C^{20}=g_{\mu\nu}C^{\mu\nu}-\displaystyle\frac{1}{2}. 
\end{eqnarray*}

\end{document}